\documentclass[conference]{IEEEtran}
\IEEEoverridecommandlockouts

\usepackage{cite}
\usepackage{amsmath,amssymb,amsfonts}

\usepackage{graphicx}
\usepackage{textcomp}
\usepackage{xcolor}
\usepackage{algorithm} 
\usepackage{algpseudocode} 
\usepackage[inline]{enumitem}
\usepackage{listings}
\usepackage[switch]{lineno}
\def\BibTeX{{\rm B\kern-.05em{\sc i\kern-.025em b}\kern-.08em
    T\kern-.1667em\lower.7ex\hbox{E}\kern-.125emX}}
\begin{document}

\title{Find Unique Usages: Helping Developers Understand Common Usages}

\author{
\IEEEauthorblockN{Emad Aghayi}
\IEEEauthorblockA{\textit{Department of Computer Science} \\
\textit{George Mason University}\\
Fairfax, VA \\
eaghayi@gmu.edu}
\and
\IEEEauthorblockN{Aaron Massey}
\IEEEauthorblockA{\textit{Department of Computer Science} \\
\textit{George Mason University}\\
Fairfax, VA \\
amassey5@gmu.edu}
\and
\IEEEauthorblockN{Thomas D. LaToza}
\IEEEauthorblockA{\textit{Department of Computer Science} \\
\textit{George Mason University}\\
Fairfax, VA \\
tlatoza@gmu.edu}
}

\maketitle

\begin{abstract}
When working in large and complex codebases, developers face challenges using \textit{Find Usages} to understand how to reuse classes and methods. To better understand these challenges, we conducted a small exploratory study with 4 participants. We found that developers often wasted time reading long lists of similar usages or prematurely focused on a single usage. Based on these findings, we hypothesized that clustering usages by the similarity of their surrounding context might enable developers to more rapidly understand how to use a function. To explore this idea, we designed and implemented \textit{Find Unique Usages}, which extracts usages, computes a diff between pairs of usages, generates similarity scores, and uses these scores to form usage clusters. To evaluate this approach, we conducted a controlled experiment with 12 participants. We found that developers with \textit{Find Unique Usages} were  significantly  faster,  completing  their  task  in  35\%  less time.
\end{abstract}

\begin{IEEEkeywords}
Software reuse, code navigation, programming tools, development environments
\end{IEEEkeywords}

\section{Introduction}

When developers look to reuse existing functionality in their codebase by interacting with classes and calling methods, developers work to understand existing code. Developers report that understanding existing code is one of their most time-consuming activities~\cite{latoza2006maintaining}. Developers generally avoid relying on documentation, which is frequently out of date and poorly written~\cite{documentation} and may not answer low-level questions about hidden contracts, implementation details, or side effects~\cite{head2018not}. Instead, developers tend to rely primarily on the code itself~\cite{head2018not, latoza2006maintaining}. The most frequent developer activity is code search~\cite{singer2010examination}, where developers most often look for code examples ~\cite{sadowski2015developers}. 
94\% of developers search when they are working on maintenance tasks~\cite{lawrance2008using}. To do these searches, developers spend a significant amount of time navigating existing code~\cite{piorkowski2016foraging,ko2006exploratory}. Given the centrality of navigating code to find and understand examples in programming tasks, even small improvements in making this process more successful and effective may have an important impact.

A variety of tools assist developers in navigating and searching code~\cite{augustine2015field,ko2006exploratory,albusays2017interviews}. Modern development environments offer tools for developers to \textit{Find Usages} or navigate the \textit{Call Hierarchy}. For example, the JetBrains IDE offers developers a window which lists results of all usages of a class or method. To understand how to use a method, a developer may simply read each invocation of a method. In this way, these IDE features are envisioned to support the process of code reuse and help developers to identify and understand the ways in which methods are used.

To investigate the challenges developers may face in using \textit{Find Usages} to reuse code, we conducted a small exploratory study where we observed four developers implementing a feature in an open source codebase. We found that developers sometimes became overwhelmed by the number of usage results listed. Instead of investigating more usages, participants instead focused on a single usage as an example for some time before moving on to investigate other potentially better examples. This would result in participants learning less from code examples through usages than they could have.
Based on these findings, we hypothesized that, rather than listing all usages separately, developers might be able to more rapidly understand how to use a function through a usages view organized into groups based on the similarity of the surrounding code. \par

To explore this idea, we propose \textit{Find Unique Usages}, which clusters usages by similarity and displays usage clusters to the developer. \textit{Find Unique Usages} extracts usages of elements, computes a diff between pairs of usages, generates similarity scores, and uses these scores to form usage clusters.
To evaluate \textit{Find Unique Usages}, we conducted  an experiment where 12 developers implemented a small feature in an open-source codebase. We found that developers with \textit{Find Unique Usages} were significantly faster, completing their task in 35\% less time.\par

In the rest of this paper, we first review related work. We then describe our exploratory study of how developers use \textit{Find Usages}. Based on our findings, we then present the design of \textit{Find Unique Usages} and an evaluation of its use. Finally, we conclude with a discussion of limitations as well as opportunities and future directions.

\begin{figure*}
    \centering
    \includegraphics [width=\textwidth,keepaspectratio,clip]{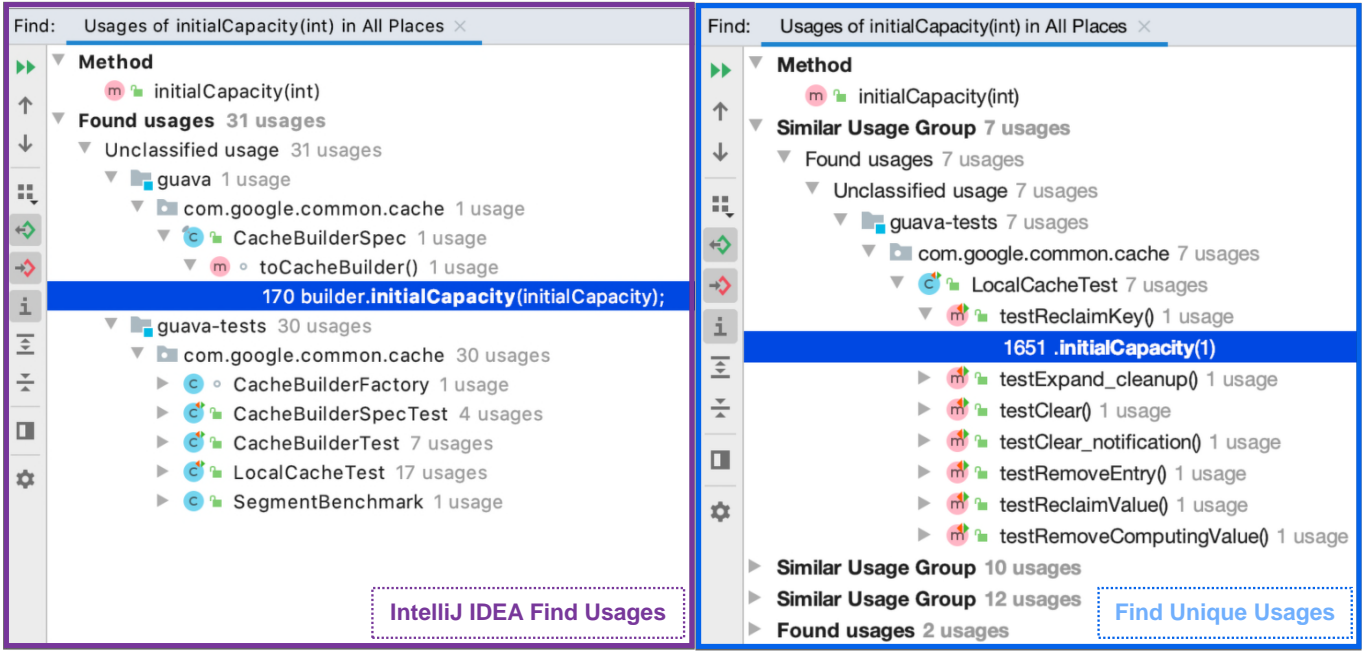}
    \caption{Using the traditional \textit{Find Usages} tool, 31 usages are listed based on the package where they are located. In \textit{Find Unique Usages}, these 31 usages are clustered into three clusters, offering developers a summarized view of usages.}
\label{fig:compare}
\end{figure*}
\section{Related Work}
Our work builds on prior studies of how developers work to understand code and tools to help developers in understanding code.

In their daily tasks, developers collect information from peers, code, documentation, and other resources~\cite{latoza2006maintaining}. Information foraging theory~\cite{pirolli1999informationforaging} has been used to describe how software developers search for information in code~\cite{fleming2013information}. According to this theory, developers try to maximize the value of information they expect to obtain and minimize the cost of navigation. When developers choose where to look, they estimate the expected value of information they find in that location and the cost of finding it. Half of navigation choices lead to less value than what developers expected~\cite{piorkowski2016foraging}. \par

Understanding how developers navigate while foraging for information is important. Developers spend 35\% to 50\% of their time navigating through source code during software development activities~\cite{ko2006exploratory,piorkowski2013whats}. Navigating and re-finding places in code that have already been visited is frequent, difficult, and distracting~\cite{ko2005eliciting,deline2005towards}. Many tools try to help developers optimize code navigation. Structural relationship traversal tools let developers traverse relationships between code elements to find other related code~\cite{karrer2011stacksplorer,augustine2015field,latoza2011visualizing}. Recommender tools predict relevant elements based on the history generated when developers worked on similar tasks~\cite{zimmermann2005mining,deline2005easing}. Task context navigation tools make it easier to navigate back and forth between task context elements~\cite{ko2006exploratory}. However, code navigation remains challenging for developers~\cite{albusays2017interviews}.\par

One reason developers forage for information is to find code examples~\cite{rosson1996reuse, brandt2009two}. Developers look for examples of how to use specific methods or objects~\cite{stylos2006mica,umarji2008archetypal}. Opportunistic developers are more likely to use example code rather than systematic developers~\cite{head2018not}. There exist a number of techniques intended to support code reuse. For instance, IDEs have search tools that enable developers to look for example code.\par

Developers often reuse existing code in their codebase to complete their tasks. A common way to edit code is by copying similar existing code, creating a code clone~\cite{codeCloneDetection2019,hou2009cnp}. Developers create code clones to apply basic templates, apply design patterns, or reuse the definition of specific behavior~\cite{kim2004ethnographic,kapser2008cloning}. Code clones in many cases are harmful, but there may be some situations where code clones are beneficial.  Many tools detect harmful code clones in codebases, identifying code duplication indicating code smells to be refactored~\cite{bellon2007comparison}. Our tool builds on this work, adapting ideas for finding duplicated code to helping developers understand similar usage sites.\par



\begin{figure*}[h]
    \centering
    \includegraphics [width=15cm,height=10cm,keepaspectratio,clip]{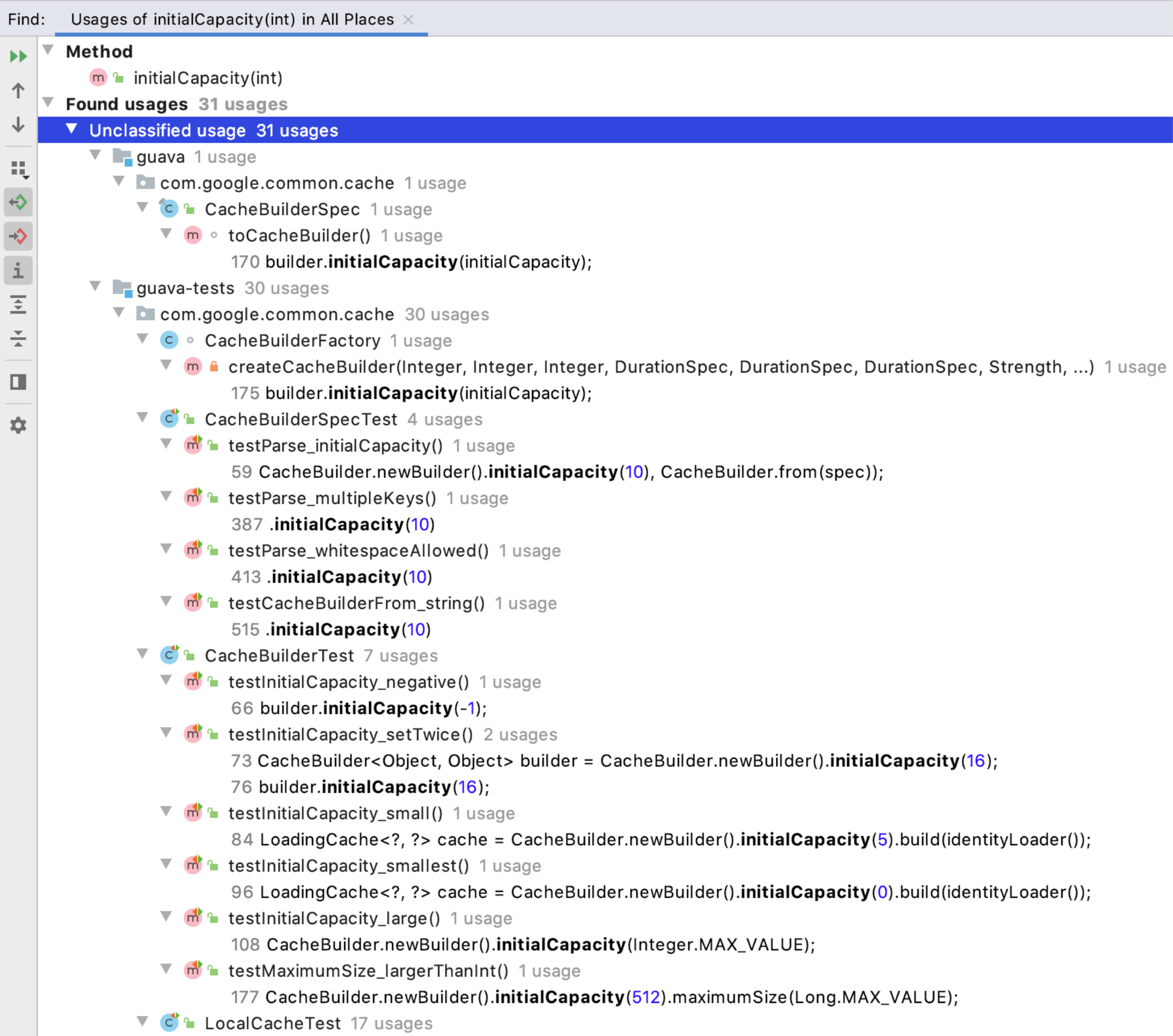}
    \caption{Developers face challenges in working with the \textit{Find Usages} tool. In this example, the results list 31 usages of the "initCapacity" method. Developers were forced to go through these 31 results. 
    While many provided the same input values (e.g., 10), discovering this was annoying and time consuming for participants. 
}
\label{fig:usege}
\end{figure*}
\section{Study 1: Challenges with Find Usages}

To better understand how developers use \textit{Find Usages} and the challenges they face, we conducted a formative observational study in which 4 participants worked to implement a feature in an unfamiliar codebase.

\subsection{Method}
We recruited four participants (P1, P2, P3, P4), two female and two male. The first participant (P1) was a graduate computer science student at our institution and participated as a pilot participant. The remaining three participants worked as a software developer in industry. Two participants had less than a year of industry experience (P2 and P4), and the third had more than four years of experience (P3).\par 


To ensure participants were familiar with the \textit{Find Usages} tools, participants first completed a training task. Three of the four participants were unaware of \textit{Find Usages}. 
Participants worked in the Google Guava project, a 772,475 LOC open-source library written in Java. To focus participants on familiarizing themselves with \textit{Find Usages}, we asked participants to qualitatively describe the range of integer inputs to two methods used in the codebase. Participants were given a maximum of 10 minutes to reach the point where they had become comfortable with the \textit{Find Usages} tool. \par

To observe developers working with \textit{Find Usages} in a more realistic programming task, the main task asked participants to implement a feature. Participants worked on the FlyingSaucer project, a pure Java library for rendering XML, XHTML, and CSS written in approximately 99,000 LOC. We removed the statements that create a PDF and asked participants to implement functionality to produce “success.pdf.” To focus participants' attention on the code, participants were instructed to treat the codebase as closed source. They were instructed not to look for online documentation or code examples. Participants were free to choose any IDE they wished to use to accomplish the task.

During the study, participants first completed the training task before beginning the main task. 
Participants were asked to think-aloud as they worked.
At the and of the study, participants completed a survey about their experiences. We asked participants to share challenges they experienced 1) finding a method, variable, or other element, 2) beginning work, 3) understanding and working with the provided codebase, 4) related to their productivity. The study lasted approximately 50 minutes for each participant. 

We piloted our initial study design with the graduate student participant (P1). We found the tasks to be appropriate and clarified the instructions. The other participants reported that the tasks were similar to what they did in their job in industry. 
\begin{quote} "This is just like my job" and "I do this at work all the time, no one knows how anything works but we see how things are used"
- (P3)
\end{quote}

\begin{figure*}
    \centering
    \includegraphics [width=\textwidth,keepaspectratio,clip]{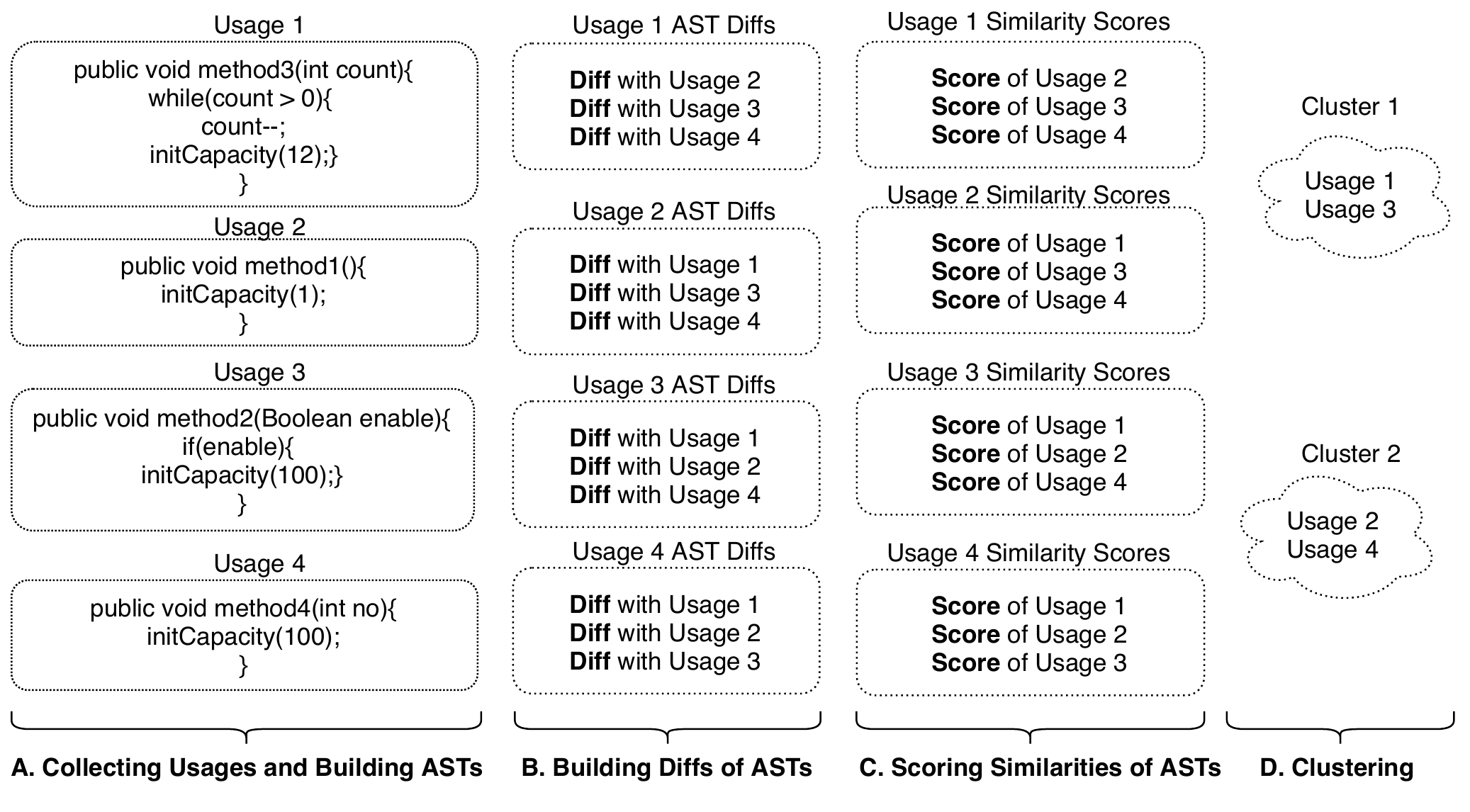}
    \caption{\textit{Find Unique Usages} computes usage clusters through four steps. It first collects usages and creates an AST for each usage. It then builds a diff between each pair of usages. From these diffs, it then calculates a similarity score between each pair. Finally, it clusters usages based on their similarity scores, using a threshold to determine when to assign it to an existing cluster or create a new cluster.}
\label{fig:generalview}
\end{figure*}

\subsection{Results}

All three participants successfully completed the main task, working for 49, 20, 42 minutes. We examined their activity to understand their use of \textit{Find Usages} and the challenges they faced. 

Participants were often exploratory in their work, making decisions without exhaustively gathering evidence. 
\begin{quote}"I do not know how .layout() is used but I am going to use it because it's in this code" - (P2) \end{quote}
\begin{quote}"I feel like this [.layout()] should just work, I hope!" - (P4) \end{quote}

\noindent Two participants felt overwhelmed by the size of the codebase.
\begin{quote}"Overwhelmed by the amount of code." - (P2) \end{quote}
\begin{quote}"Large codebases are my biggest fear." - (P4) \end{quote}

Two participants used the unit tests included in the project as examples to determine how methods are typically invoked and what the possible input arguments for those methods may be. Participants often stuck with a specific test case, and did not go looking for others. 
\begin{quote} "Tests are a good example of uses." - (P3)\end{quote}

The similarity of \textit{Find Usages} results was a significant challenge for participants. When many results were found (e.g., Fig.~\ref{fig:usege}), it was challenging for participants to identify useful information in understanding the method or object. 
Much of the code looked similar, and it was difficult for the participant to observe differences. This was particularly challenging when usages were complex, and they navigated call graphs.
\begin{quote}"There were a ton of methods and usages that were really similar and it was a lot to put together"- (P4)\end{quote}
\begin{quote}"I had difficulty finding usages with low complexity of calls and uses"- (P4)\end{quote}

Participants often scrolled quickly through usages when the surrounding code did not make calls similar to those they wished to make. All participants sometimes navigated to usages that did not help them complete the task and wasted time. 
Surprisingly, participants tended to focus on the first usage. By default, IntelliJ IDEA highlights the first usage in the results window. Participants copied the first usage, pasted it where this wished to reuse the functionality, and adapted it as they needed.\par 

One participant had difficulty working with \textit{Find Usages} in the presence of overloaded methods. When they found usages where the types or number of input arguments were different, they navigated between them to understand why there were different implementations of a method.


\section{Find Unique Usages Tool}
The formative study revealed that developers face challenges working with Find Usage results containing many highly similar results. We hypothesized that clustering similar usages might help developers understand usages more quickly and easily. 
To explore this hypothesis, we designed \textit{Find Unique Usages} (Fig.\ref{fig:compare}). 
\textit{Find Unique Usages} clusters usages based on each usage's surrounding code and displays the clusters to the user. To identify similar usages, we use  Gumtree Spoon AST Diff~\cite{falleri2014fine} to calculate a diff between each usage site and then iteratively cluster usages. Fig.~\ref{fig:generalview} offers an overview of our approach. \textit{Find Unique Usages} computes usage clusters through four steps: 
1) collecting raw usages and creating ASTs for each usage, 2) calculating a diff between the ASTs of each pair of usages, 3) using the diffs to calculate a similarity score for each pair of usages, 4) iteratively clustering usages based on the similarity scores. \textit{Find Unique Usages} is implemented as an IntelliJ plugin.





\subsection{Collecting Usages and Building ASTs}
In the first step, \textit{Find Unique Usages} collects usages from the codebase. For each usage, it generates an AST, which includes the entire body of the method containing the usage. The list of usage statements is retrieved through the IntelliJ API. IntelliJ is then used to generate an AST for each usage. 
\par
 
In our initial design, we considered usages as only the usage statement itself and did not consider any of the code surrounding the usage. In preliminary tests of our tool, we found that the results found remained similar to \textit{Find Usages}. We thus increased the scope of the usage to the entire method body containing the usage. 
\par




\subsection{Building Diffs of ASTs}

Using the ASTs of each usage, \textit{Find Unique Usages} computes a diff of each pair of usages. In an early prototype, we computed the difference between ASTs using string edit distance, which we found worked poorly. We instead adopted the GumTree algorithm~\cite{baxter1998clone,DBLP:conf/kbse/FalleriMBMM14,falleri2014fine}, which computes a structure aware diff of each pair of function bodies. GumTree was initially created to support Source Code Management (SCM), while we adopt it to cluster code. \par

\subsection{Scoring Similarity}
Rather than identify usages which constitute an exact match or count only the number of nodes which vary in the diff, we adapted an approach from prior work for computing similarity~\cite{baxter1998clone}. 
Similarity is computed as:
\begin{equation}
Similarity = 2 \times Shared  \div (2  \times Shared  + AST1 + AST2)
\label{equation1}
\end{equation}

\noindent where $Shared$ is the number of shared nodes between two trees calculated by GumTree, $AST1$ is the number of nodes which differ in usage  1 and $AST2$ is the number of nodes which differ in usage 2.
Similarity scores are calculated for all pairs of usages. 

\subsection{Clustering Usages}
\textit{Find Unique Usages} next computes clusters of usages based on their similarity scores. It use a max of min algorithm. It first finds the minimum similarity between the usage and all members of all clusters separately and memoizes them. Based on max of mins similarity, it chooses the best cluster. To do so, it uses two algorithms, one to find the minimum similarity of all clusters with the usage and another to find the best cluster.\par 

Algorithm 1 first finds a minimum similarity between a usage to be clustered and all members of a cluster.\par

\begin{algorithm}
\label{algo1}
    \caption{Minimum Similarity in a Usage cluster - minSimilarity($x$, $G_{i}$)} 
    \begin{algorithmic}[1]
    \State Given a Usage $x$ and a usage cluster $G_{i}$
    \State $minSimilarity$ $\leftarrow$ $\infty$
    \For{each usage $u_{i}$ in $G_{i}$}
    \State $minSimilarity$$\leftarrow$min($minSimilarity$,similarity($x$,$u_{i}$)) \EndFor
   \State retrun $minSimilarity$
    \end{algorithmic} 
     \
\end{algorithm}

Next, Algorithm 2 is used to find the most similar cluster for the usage. This algorithm iterates over all clusters, using Algorithm 1 to find the most similar cluster. It next considers a similarity \textit{threshold} to determine if the cluster is sufficiently threshold. The threshold controls the number of clusters which will be created, with higher thresholds generating fewer clusters~\cite{deng2013top}. 
We arbitrarily chose a similarity threshold of approximately 88\% after experimenting with a number of examples. The algorithm then uses this threshold to determine if a usage should be placed in a new cluster or added to the most similar cluster. 

\begin{algorithm}
\label{algo2}
    \caption{Find Corresponding Usage cluster} 
    \begin{algorithmic}[1]
    \State // Given a Usage $x$ \& a set of Usage Clusters $G$
    \State $mostSimilarCluster$ $\leftarrow$ $null$
    \For{each Usage Cluster $G_{i}$ in $G$}
    \If{minSimilarity($x$,$mostSimilarCluster$)$<$ minSimilarity($x$, $G_{i}$)}
    \State $mostSimilarCluster$ $\leftarrow$ $G_{i}$
    \EndIf
    \EndFor
    \State // Given some similarity threshold $T$
    \If{minSimilarity($x$, $mostSimilarCluster$) $<$ $T$}
        \State // Create new usage cluster and modify $G$.
        \State $G_{new}$ $\leftarrow$ newUsageClusterWithInitialMember($x$)
        \State $G$ $\leftarrow$ $G$ $\cup$ \{$G_{new}$\}
    \Else
    \State addToCluster($mostSimilarCluster$, $x$)
    \EndIf
    \end{algorithmic} 
\end{algorithm}

\subsection{User Interface}
As adopting unfamiliar tools may impose an additional burden on developers~\cite{adaption2002}, we implemented \textit{Find Unique Usages} by extending the existing \textit{Find Usages} interface in the IntelliJ IDEA (Fig. \ref{fig:compare}). 
Our interface reorganizes the list view, adding usage clusters containing usages with similar code. 
One key design choice was whether to present results in an inline pane or separate window. We chose to present usages in a separate window to enable more detailed presentation of results. \par

\section{Study 2: Evaluation}
To evaluate \textit{Find Unique Usages}, we conducted a between subjects controlled experiment in which 12 participants worked to add a feature to a codebase. 
\subsection{Method}
We recruited 13 participants by advertising on social networks. One participant with insufficient Java experience left the study early. We excluded this participant, yielding 12 participants in our study, C1 to C6 and E1 to E6. 
33\% of participants were female, and 66\% male. Five participants were graduate students, four worked as a software developers, and 3 were undergraduates. Participants in both conditions had comparable levels of Java experience, with 60 months for control and 62 for experimental participants. All participants were volunteers.\par

Participants were randomly assigned to a control  or experimental condition. Two of the undergraduates were in the experimental condition, and one was in the control condition.  All participants used the IntelliJ IDEA. Control participants used the regular \textit{Find Usages} tool and experimental participants used \textit{Find Unique Usages}.
Participants completed the same tasks as tasks in the exploratory study.
At the end of the study, we conducted a semi-structured interview and asked participants about their experiences using the \textit{Find Unique Usages} and \textit{Find Usages} tools.

\subsection{Results}
\subsubsection{Task performance}
All participants in both conditions successfully completed the task. Experimental participants completed the task in significantly less time (t = 1.82, p $<$ 0.049), 
finishing in 21.5 minutes (SD = 7.65) compared to 33 minutes for control participants (SD = 9.35). The effect size for Glass's delta is 1.24
and for Cohen's d = 1.36
. This indicates that the mean of the experimental group is at the $90_{th}$ percentile of the control group.  



\subsubsection{Interacting with lists of usages} All participants in both groups read lists of usages sequentially. They began from the first result in the list and proceeded further.
The \textit{Find Usages} view of IntelliJ IDE supports this behavior by expanding and highlighting the first usage. 
However, none of the participants read all of the usage results. 
More successful participants used a strategy in which they first used \textit{Find Usages} with the \textit{Find In Path} tools for understanding the codebase. Before they started reading usages, they expanded and skimmed the list of usages. From this, they selected the best usage that might help them. \par

Usages listed in the results window were easier to read when they contained literals directly in the call site rather than referencing variables or expressions defined elsewhere. When they did contain variables, developers were forced to open the class containing the usage to read and understand it. \par

\subsubsection{Navigating between the list of usages and code}Almost all participants cycled through the following steps.
Participants first clicked on a usage that had an object as an input argument. Next, they went to the class defining the object. They scrolled over the class's methods and read the code of the class. Some, but not all, tried to understand that class and its methods by invoking \textit{Find Usages}. Several participants then reported being lost and experienced difficulty understanding the class. In either case, participants then returned back to the first usage. As they had invoked another Find Usage command, they spent time remembering where they were and re-invoking the first command they began with. In this way, participants cycled through the steps of 
selecting a usage, navigating to the class defining the usage, invoking another find usage to understand usages of this code, reading code, and returning back to the starting point. \par

All participants experienced challenges making recursive use of \textit{Find Usages}. Participants selected a usage, opened the class containing that usage, and again invoked \textit{Find Usages} on other methods. After invoking \textit{Find Usages} several times, they sometimes lost their place in the call graph and became disoriented. \par

In both conditions, four participants struggled with information overload. For example, in their task, they needed to use a method that was overloaded. This lead to methods that had the same name, but had a different number of input parameters, types of input parameters, or both. When participants were using \textit{Find Usages}, this confused them.\par

\subsubsection{Usability issues}
At the end of the study, we interviewed  participants about their experiences. Participants reported a number of usability issues. One participant reported that recursive use of \textit{Find Usages} was confusing. 
\begin{quote} "I was getting lost when I was using nested [several sequential] \textit{Find Usages} for understand codebase." - (C6)\end{quote}
\noindent C6 and E3 experienced a usability issue with the line number of the usage statement in the file, which they instead interpreted as the frequency.
E2 and E4 found the use of the same name for all clusters in \textit{Find Unique Usages} to be confusing (see Fig.~\ref{fig:compare}). 
C6 specifically requested support for combining \textit{Find Usages} with call graph navigation to better support recursively investigating usages. During the study, he got lost and felt that a call graph navigation tool would help him stay oriented more easily.\par


IntelliJ IDEA offers an inline list of \textit{Find Usages}, invoked by pressing control on the keyboard and clicking on a method. In this approach, developers do not go to a separate results window. The two participants that discovered this feature, one each from the control and experimental group, completed the task in the two shortest times (17 and 13 minutes).



\section{Limitations And Threats To Validity}
Like all studies, our study has several important limitations and potential threats to validity. 
Unlike most developers, our participants had no experience with the codebase in which they worked. Developers with more knowledge might navigate less and rely more on their existing code knowledge, resulting in different interactions with \textit{Find Usages}. 
Developers also did not benefit from documentation or the ability to ask coworkers questions, which might have changed their information needs and the strategies they used to satisfy them. 
To ensure participants were familiar with \textit{Find Usages}, we trained them in its use, enabling a comparison between use of \textit{Find Usages} and \textit{Find Unique Usages}. But participants may not have been familiar with other alternative navigational aids, which they might have instead used. In choosing the codebase in which participants worked, we sought to identify a representative Java codebase of medium size. But codebases with usage sites which differed, such as by containing more or fewer literals or overloaded methods, might have led to different results.

\section{Discussion}
Our exploratory study demonstrated that developers can face significant challenges in using \textit{Find Usages} to understand how to reuse a method or class. Developers faced challenges with the high number of similar usages, often focusing on the first usage without considering others. To help developers more easily survey the range of usages, we developed a new technique to identify unique usages. Our approach utilizes AST diffs of usage contexts to cluster similar usages. We found that developers with \textit{Find Unique Usages} were able to complete a task implementing simple logic in an unfamiliar codebase in 35\% less time.

Only one participant surveyed all usages before beginning reading usages. By default, both \textit{Find Unique Usages} and \textit{Find Usages} in IntelliJ expand and highlight the first usage in the list. This may be one reason many developers choose to focus on the first usage. It is unclear how developers' behavior might change if the IDE did not highlight this first usage. An alternative design might be to expand all usages by default and not highlight any usage. 
This design is similar to the inline \textit{Find Usages} offered, which provides only a list of usages and hides other complexity. \par


\textit{Find Unique Usages} is highly sensitive to the threshold selected, which determines the number of clusters created. Future work might more systematically investigate the impact of the number of clusters chosen. The developer might also be given more direct control, helping them to see an overview of clusters and drill into interesting clusters in more detail.
More broadly, many questions remain about just what makes a usage distinct for a developer and how developers might wish to understand these usage clusters.\par


There are a wide range of clustering techniques that might be used to cluster usage sites. For example, hierarchical clustering is a common clustering approach for collecting item into groups based on similarity or distance scores.  Hierarchical clustering commonly involves merging groups, which we did not implement in our approach~\cite{shanmughasundaram2015measurement,day1984efficient,davidson2005agglomerative}. \par

Our results offer additional evidence for the value of call graph navigation tools. Rather than simply looking at usages one level deep, developers often wished to understand usages by going deeper. A wide variety of call graph navigation tools have been designed to support this behavior, beginning with tools such as the call hierarchy and extending to more sophisticated research tools such as StackSplorer~\cite{karrer2011stacksplorer}, Reacher~\cite{latoza2011visualizing}, and Prodet~\cite{augustine2015field}. Our results suggest that these tools may help support developers in understanding method usages. To more directly help developers both understand which usages are distinct and worth understanding as well as understand usages, it might be possible to create new hybrid tools. For example, a tool might combine a view showing clustered usages with the ability to understand the usage in more detail by following paths through the call graph. As distinct usages involve more than one code example, it may also be valuable to find ways to summarize multiple similar paths.

\section*{Acknowledgments}
We thank Jonathan Bell for his contributions in helping brainstorm the idea of this research, we thank the participants in the study for their participation, and we also thank those enrolled in SWE 795: Software Engineering Environments in Fall 2019.
This research was supported in part by the National Science Foundation under grants CCF-1414197 and CCF-1845508.

\bibliographystyle{IEEEtran}
\bibliography{FUU}

\end{document}